\documentclass[12pt]{article}
\usepackage{amsmath}
\usepackage{amssymb}
\usepackage{graphicx}
\usepackage{epsfig}
\usepackage{cite}
\usepackage{url}
\usepackage{array,arydshln}
\usepackage[small]{caption}
\begin{document}
\baselineskip 0.6cm

\def\simgt{\mathrel{\lower2.5pt\vbox{\lineskip=0pt\baselineskip=0pt
           \hbox{$>$}\hbox{$\sim$}}}}
\def\simlt{\mathrel{\lower2.5pt\vbox{\lineskip=0pt\baselineskip=0pt
           \hbox{$<$}\hbox{$\sim$}}}}
\def\simprop{\mathrel{\lower3.0pt\vbox{\lineskip=1.0pt\baselineskip=0pt
             \hbox{$\propto$}\hbox{$\sim$}}}}
\def\bra#1{\langle #1 |}
\def\ket#1{| #1 \rangle}
\def\inner#1#2{\left< #1 | #2 \right>}

\begin{titlepage}

\begin{flushright}
UCB-PTH-13/07\\
\end{flushright}

\vskip 1.5cm

\begin{center}
{\Large \bf Black Holes or Firewalls: A Theory of Horizons}

\vskip 0.7cm

{\large Yasunori Nomura, Jaime Varela, and Sean J. Weinberg}

\vskip 0.4cm

{\it Berkeley Center for Theoretical Physics, Department of Physics,\\
 University of California, Berkeley, CA 94720, USA}

\vskip 0.1cm

{\it Theoretical Physics Group, Lawrence Berkeley National Laboratory,
 CA 94720, USA}

\vskip 0.8cm

\abstract{We present a quantum theory of black hole (and other) horizons, 
 in which the standard assumptions of complementarity are preserved 
 without contradicting information theoretic considerations.  After 
 the scrambling time, the quantum mechanical structure of a black hole 
 becomes that of an eternal black hole at the microscopic level.  In 
 particular, the stretched horizon degrees of freedom and the states 
 entangled with them can be mapped into the near-horizon modes in the 
 two exterior regions of an eternal black hole, whose mass is taken 
 to be that of the evolving black hole at each moment.  Salient features 
 arising from this picture include:\ (i) the number of degrees of 
 freedom needed to describe a black hole is $e^{{\cal A}/2 l_{\rm P}^2}$, 
 where ${\cal A}$ is the area of the horizon; (ii) black hole states 
 having smooth horizons, however, span only an $e^{{\cal A}/4 
 l_{\rm P}^2}$-dimensional subspace of the relevant $e^{{\cal A}/2 
 l_{\rm P}^2}$-dimensional Hilbert space; (iii) internal dynamics of 
 the horizon is such that an infalling observer finds a smooth horizon 
 with a probability of $1$ if a state stays in this subspace.  We 
 identify the structure of local operators responsible for describing 
 semi-classical physics in the exterior and interior spacetime regions, 
 and show that this structure avoids the arguments for firewalls---the 
 horizon can keep being smooth throughout the evolution.  We discuss 
 the fate of infalling observers under various circumstances, especially 
 when the observers manipulate degrees of freedom before entering the 
 horizon, and we find that an observer can never see a firewall by 
 making a measurement on early Hawking radiation.  We also consider the 
 presented framework from the viewpoint of an infalling reference frame, 
 and argue that Minkowski-like vacua are not unique.  In particular, 
 the number of true Minkowski vacua is infinite, although the label 
 discriminating these vacua cannot be accessed in usual non-gravitational 
 quantum field theory.  An application of the framework to de~Sitter 
 horizons is also discussed.}

\end{center}
\end{titlepage}

\section{Introduction and Summary}
\label{sec:intro}

General relativity and quantum mechanics are two pillars in contemporary 
fundamental physics.  The relation between the two, however, is not clear. 
On one hand, one can build quantum field theory on a fixed curved background, 
calculating quantum properties of matter in the existence of gravity. 
On the other hand, a naive application of such a semi-classical procedure 
often leads to puzzles that signal the incompleteness of the picture. 
A well-known example is the overcounting of degrees of freedom that arises 
when the interior spacetime and outgoing Hawking radiation of a black hole 
are treated as independent objects on a certain equal-time hypersurface 
(called a nice slice)~\cite{Preskill:1992tc}.  It is clearly an important 
and nontrivial task to understand how the world as described by general 
relativity emerges in a consistent theory of quantum gravity.

An elegant way to address the overcounting problem described above was 
put forward in Refs.~\cite{Susskind:1993if,Stephens:1993an} under the 
name of black hole complementarity.  This hypothesis asserts that
\begin{itemize}
\item[(i)]
The formation and evaporation of a black hole are described by unitary 
quantum evolution.
\item[(ii)]
The region outside the stretched horizon is well described by quantum 
field theory in curved spacetime.
\item[(iii)]
The number of quantum mechanical degrees of freedom associated with 
the black hole, when described by a distant observer, is given by the 
Bekenstein-Hawking entropy~\cite{Bekenstein:1973ur}.
\item[(iv)]
An infalling observer does not feel anything special at the horizon 
(no drama) consistently with the equivalence principle.
\end{itemize}
With these assumptions, the issue of overcounting can be solved---the 
distant picture having Hawking radiation and the infalling picture 
with the interior spacetime are two different descriptions of the 
same physics; in particular, they are related by a unitary transformation 
associated with the reference frame change~\cite{Nomura:2011rb}. 
This complementarity picture, however, has recently been challenged 
in Refs.~\cite{Almheiri:2012rt,Almheiri:2013hfa,Marolf:2013dba,%
Braunstein:2009my}, which assert that the smoothness of horizon as 
implied by general relativity, (iv), is incompatible with the other 
assumptions, (i)~--~(iii).  If true, this would have profound implications 
for fundamental physics; in particular, it would force us to abandon 
one of the standard assumptions in contemporary physics---unitary 
quantum mechanics, locality at long distances, or the equivalence 
principle.  The authors of Refs.~\cite{Almheiri:2012rt,Almheiri:2013hfa,%
Marolf:2013dba} argue that the simplest option is to abandon the 
equivalence principle---an observer falling into a black hole hits 
a ``firewall'' of high energy quanta at the horizon.  This would 
be a dramatic deviation from the prediction of general relativity.

In this paper, we present a quantum theory of black hole (and other) 
horizons in which the standard assumptions of complementarity, (i)~--~(iv), 
are preserved.  Our construction builds on earlier observations in 
Refs.~\cite{Nomura:2012ex,Nomura,Nomura:2013nya,Verlinde:2013uja,%
Verlinde:2013vja}.  In Refs.~\cite{Nomura:2012ex,Nomura}, two of the 
authors suggested that there are exponentially many black hole vacuum 
states corresponding to the same semi-classical black hole, and that 
there can be a (semi-)classical world built on each of them, all 
of which look identical at the level of general relativity but are 
represented differently at the microscopic level.  It was argued 
that this structure can evade the firewall argument with appropriate 
internal dynamics for the horizon.  In Ref.~\cite{Nomura:2013nya}, the 
same picture was considered in an infalling reference frame in which 
the manifestation of the exponentially many microscopic states in this 
reference frame was discussed.  More recently, Verlinde and Verlinde 
considered a similar picture in which the Hilbert space structure for 
the relevant degrees of freedom was identified more explicitly and in 
which a concrete qubit model demonstrating the basic dynamics of black 
hole evaporation was presented~\cite{Verlinde:2013uja,Verlinde:2013vja}.
In this paper we develop these observations further, identifying how 
the distant and infalling descriptions as suggested by general relativity 
emerge dynamically from a full quantum state obeying the unitary evolution 
law of quantum mechanics.  In particular, we identify the structure of 
operators responsible for describing the exterior and interior regions 
of the black hole, which allows us to address explicitly the arguments 
made in Refs.~\cite{Almheiri:2012rt,Almheiri:2013hfa,Marolf:2013dba}.

\begin{figure}[t]
\begin{center}
  \includegraphics[width=16cm]{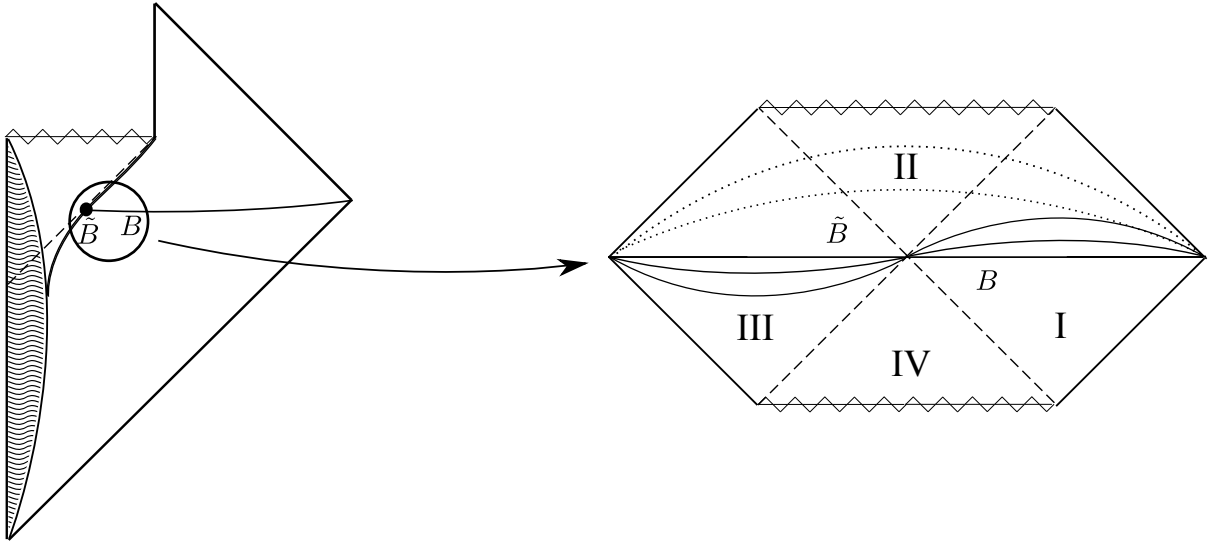}
\caption{The stretched horizon degrees of freedom, $\tilde{B}$, and the 
 states entangled with them, $B$, of an evolving black hole (left panel) 
 can be mapped into the near-horizon degrees of freedom of an eternal 
 black hole in the regions~III and I, respectively (right panel).  The 
 mapping must be made at an instant of time, with the mass of the eternal 
 black hole taken to be that of the evolving black hole at that moment. 
 The near-horizon states of the eternal black hole are defined on an 
 equal-time hypersurface determined by the outside timelike Killing 
 vector (one of the solid lines depicted).  The dotted lines in the 
 right panel indicate a succession of hypersurfaces used to obtain 
 local operators representing the interior spacetime.}
\label{fig:map}
\end{center}
\end{figure}

The basic hypothesis of our framework is:
\begin{itemize}
\item[]
The quantum mechanical structure of a black hole after the horizon is 
stabilized to a generic state (after the scrambling time~\cite{Hayden:2007cs}) 
is the same as that of an eternal black hole of the same mass at the 
microscopic level.  In particular, the degrees of freedom associated 
with the stretched horizon and the outside states entangled with them can 
be mapped to the near-horizon states of the eternal black hole in one and 
the other external regions, respectively.  (These near-horizon states are 
described in a distant reference frame, using an equal-time hypersurface 
determined by the outside timelike Killing vector.)  The precise mapping 
is such that the outside states entangled with the stretched horizon 
and the near-horizon states in one side of the eternal black hole respond 
in the same way to local operators representing physics in the exterior 
of the black hole.
\end{itemize}
It is important that this identification mapping is made in each instant 
of time; for example, the mass of the corresponding eternal black hole 
must be taken as that of the evolving black hole at each moment.  Note 
also that the identification with the eternal black hole is made only 
for the stretched horizon degrees of freedom and the outside states 
entangled with them; the structure of the other modes need not follow 
that of the eternal black hole.  We can summarize these concepts by saying 
that an eternal black hole (of a fixed mass) provides a model for an evolving 
black hole for a timescale much shorter than that of the evolution. 
A schematic picture for this mapping is depicted in Fig.~\ref{fig:map}.

Key elements to understand physics of black holes (and firewalls) arising 
from the picture described above are
\begin{itemize}
\item
The dimensions of the Hilbert spaces for the stretched horizon states 
and the states entangled with them are both $e^{{\cal A}/4 l_{\rm P}^2}$, 
where ${\cal A}$ is the area of the horizon and $l_{\rm P} \simeq 
1.62 \times 10^{-35}~{\rm m}$ is the Planck length.  The number of 
microscopic degrees of freedom needed to describe a black hole is 
thus $e^{{\cal A}/4 l_{\rm P}^2} \times e^{{\cal A}/4 l_{\rm P}^2} 
= e^{{\cal A}/2 l_{\rm P}^2}$.  The actual black hole states, however, 
occupy only a tiny $e^{{\cal A}/4 l_{\rm P}^2}$-dimensional subspace 
of the $e^{{\cal A}/2 l_{\rm P}^2}$-dimensional Hilbert space relevant 
for these degrees of freedom~\cite{Verlinde:2013uja}, as suggested by 
black hole thermodynamics.  All the other states represent ``firewall 
states,'' which do not allow for a semi-classical interpretation of 
the interior region.
\item
As long as the quantum state for the stretched horizon and the entangled 
modes stays in the $e^{{\cal A}/4 l_{\rm P}^2}$-dimensional subspace, 
an infalling observer interacting with this state finds a smooth 
horizon with a probability of $1$.  This is because the $e^{{\cal A}/4 
l_{\rm P}^2}$-dimensional subspace is spanned by $e^{{\cal A}/4 
l_{\rm P}^2}$ microstates all representing the same semi-classical 
black hole with a smooth horizon, and the internal dynamics of the 
horizon is such that an infalling object sees/measures the horizon 
in this basis~\cite{Nomura:2012ex}.  The evolution of a black hole 
is consistent with the assumption that the relevant degrees of freedom 
keep staying in this subspace so that no firewall develops.
\item
Operators responsible for describing the exterior spacetime region act 
only on the modes outside the stretched horizon as implied by local quantum 
field theory applicable outside the stretched horizon.  On the other hand, 
local operators responsible for describing the interior spacetime region 
act nontrivially both on the stretched horizon and the outside entangled 
modes.  This ``asymmetry'' arises because the stretched horizon degrees 
of freedom represent the {\it exterior} modes outside the horizon in 
the other side of the eternal black hole under the identification map 
described above.
\end{itemize}
We find that these elements elegantly address the questions raised by 
the firewall argument.  Representative results include
\begin{itemize}
\item
The evolution of a black hole does not dynamically develop a firewall 
(even after the Page time~\cite{Page:1993wv}).  An infalling observer 
who does not perform a special manipulation to his/her environment always 
sees a smooth horizon.
\item
It is not possible for an observer to see a firewall even if he/she 
performs a very special measurement on Hawking radiation emitted earlier 
from the black hole.  Such a measurement cannot change the fact that 
he/she will see a smooth horizon.
\item
If a falling observer can directly measure a mode entangled with the 
stretched horizon as he/she falls through the horizon, then he/she 
may see a firewall.  This, however, does not violate the equivalence 
principle; the same can occur at any surface in a low curvature region.
\end{itemize}
We note that the framework presented here and resulting physical 
predictions also apply to other horizons, including de~Sitter and 
Rindler horizons, with straightforward adaptations.  We will discuss 
these cases toward the end of the paper.

The organization of the rest of this paper is as follows.  In 
Section~\ref{sec:BHs}, we describe the microscopic structure of the 
black hole vacuum states.  In Section~\ref{sec:basis}, we see how these 
states are embedded in the larger Hilbert space relevant for the stretched 
horizon degrees of freedom and the states entangled with them.  We discuss 
how the vacuum and non-vacuum black hole states as well as the firewall states 
arise in this large Hilbert space, and identify the form of local operators 
responsible for describing the exterior and interior spacetime regions.  We 
argue that the dynamics of quantum gravity can be such that a black hole stays 
as a black hole state under time evolution (not becoming a firewall state), 
and that an infalling observer interacting with such a state will see 
a smooth horizon with a probability of $1$ because of the properties 
of the internal dynamics of the horizon.  In Section~\ref{sec:infalling}, 
we discuss the fate of infalling observers under various circumstances, 
especially when the observers manipulate degrees of freedom before 
entering the horizon.  We also describe how the present framework is 
realized in an infalling reference frame.  We argue that locally (and 
global) Minkowski vacuum states are not unique at the microscopic 
level, although the same semi-classical physics can be built on any one 
of them, so that this degeneracy need not be taken into account explicitly 
in usual applications of quantum field theory, e.g.\ to the problem of 
scattering.  In Section~\ref{sec:de-Sitter}, we discuss how our framework 
is applied to de~Sitter horizons.

\section{Microscopic Structure of Black Holes}
\label{sec:BHs}

Here we discuss the microscopic structure of black holes, following 
Refs.~\cite{Nomura:2012ex,Nomura,Nomura:2013nya,Verlinde:2013uja,%
Verlinde:2013vja}.  Suppose we describe a system with a black hole, which 
for simplicity we take to be a Schwarzschild black hole in 4-dimensional 
spacetime, from a distant reference frame.  We assume that, for any fixed 
black hole mass $M$, the entire system is decomposed into three subsystems:%
\footnote{We adopt a notation close to that of Ref.~\cite{Almheiri:2013hfa}, 
 although the precise physical object each symbol represents sometimes 
 differs.}
\begin{itemize}
\item[$\tilde{B}$:]
the degrees of freedom associated with the stretched horizon;
\item[$C$:]
the degrees of freedom associated with the spacetime region close to, 
but outside, the stretched horizon, e.g.\ $r \simlt 3M l_{\rm P}^2$;
\item[$R$:]
the rest of the system (which may contain Hawking radiation emitted 
earlier).
\end{itemize}
Among all the possible quantum states for the $C$ degrees of freedom, 
some are strongly entangled with the states representing $\tilde{B}$.%
\footnote{We ignore possible tiny direct entanglement between $\tilde{B}$ 
 and $R$, which is a good approximation if the spacetime region for $C$ 
 is taken sufficiently large.  Note that the entanglement necessary for 
 unitarity of the evolution of the state is not of this type; see later.}
We call the set of these quantum states $B$:
\begin{itemize}
\item[$B$:]
the quantum states representing the states for the $C$ degrees of freedom 
that are strongly entangled with the degrees of freedom described by 
$\tilde{B}$.
\end{itemize}
Following the locality hypothesis, we consider that systems $C$ and $R$, 
more precisely operators acting only on $C$ or $R$, are responsible for 
physics outside the stretched horizon, which is well described by local 
quantum field theory at length scales larger than the fundamental (string) 
length $l_*$.  On the other hand, the interior spacetime for an infalling 
observer, as we will argue, is represented by operators acting on the 
combined $\tilde{B}B$ system (on both $\tilde{B}$ and $B$ states).  In 
our analysis below, we ignore the center-of-mass drift and spontaneous 
spin-up of black holes~\cite{Page:1979tc}, which give only minor effects 
on the dynamics.

Suppose, as usual, we quantize the system in such a way that the 
Hamiltonian near (and outside) the horizon takes locally the Rindler 
form.  Then, a black hole vacuum state is described by one in which some 
of the states for the $C$ degrees of freedom, i.e.\ $B$, are (nearly) 
maximally entangled with the states for $\tilde{B}$~\cite{Unruh:1976db}. 
The basic idea of Refs.~\cite{Nomura:2012ex,Nomura} is that there are 
exponentially many ($\approx e^{{\cal A}/4 l_{\rm P}^2}$ where ${\cal A} 
= 16\pi M^2 l_{\rm P}^4$ is the horizon area%
\footnote{Here and below, similar expressions are valid at the leading 
 order in expansion in powers of $l_{\rm P}^2/{\cal A}$, in the exponent 
 for the number of states (or in entropies).  With this understanding, 
 we will use the equal sign below, instead of the approximate sign.}%
) black hole vacuum states $\ket{\psi_i}$ which correspond to the same 
semi-classical black hole, and that there can be a (semi-)classical 
world built on each of them, all of which look {\it identical} to general 
relativity but are represented differently at the microscopic level 
(consistently with the no-hair theorem).  More specifically, the states 
$\ket{\psi_i}$, which live in the combined $\tilde{B}B$ system, can be 
written as~\cite{Nomura,Verlinde:2013uja}
\begin{equation}
  \ket{\psi_i} = \sum_{j=1}^{e^{{\cal A}/4 l_{\rm P}^2}}
    \alpha^{(i)}_j \ket{\tilde{b}_j} \ket{b_j},
\qquad
  (i = 1,\cdots,e^{{\cal A}/4 l_{\rm P}^2}).
\label{eq:psi_i}
\end{equation}
Here, $\alpha^{(i)}_j$ are coefficients that satisfy the orthonormality 
condition and the condition for each $\ket{\psi_i}$ being maximally 
entangled
\begin{equation}
  \sum_{j=1}^{e^{{\cal A}/4 l_{\rm P}^2}} \alpha^{(i)*}_j \alpha^{(i')}_j 
  = \delta_{ii'},
\qquad
  |\alpha^{(i)}_j|^2 = e^{-\frac{\cal A}{4 l_{\rm P}^2}},
\label{eq:ortho-max}
\end{equation}
and $\ket{\tilde{b}_j}$ and $\ket{b_k}$ ($j,k = 1,\cdots,
e^{{\cal A}/4 l_{\rm P}^2}$) are elements of ${\cal H}_{\tilde{B}}$ 
and ${\cal H}_B$ with~\cite{Verlinde:2013uja}
\begin{equation}
  {\rm dim}\,{\cal H}_{\tilde{B}} = {\rm dim}\,{\cal H}_B 
  = e^{\frac{\cal A}{4 l_{\rm P}^2}},
\label{eq:H_Bs}
\end{equation}
where ${\cal H}_{\tilde{B}}$ and ${\cal H}_B$ are the Hilbert space 
factors that contain all the possible states for $\tilde{B}$ and $B$, 
respectively.

To be more precise, the black hole vacuum states $\ket{\psi_i}$ are 
written as
\begin{equation}
  \ket{\psi_i} = \sum_{j=1}^{j_{\rm max}} 
    e^{-\frac{\beta_j}{2}E_j} \alpha^{(i)}_j \ket{\tilde{b}_j} \ket{b_j},
\label{eq:psi_i-Boltzmann}
\end{equation}
instead of Eq.~(\ref{eq:psi_i}).  Here, $|\alpha^{(i)}_j|^2 = 
1/\sum_{j'=1}^{j'_{\rm max}} e^{-\beta_{j'} E_{j'}}$, and $E_j$ and 
$\beta_j$ are the energy of the state $\ket{b_j}$ and the reciprocal of 
the temperature relevant for it (i.e.\ the effective blue-shifted local 
Hawking temperature relevant for the state).  The expressions in 
Eqs.~(\ref{eq:psi_i}~--~\ref{eq:H_Bs}) are the ones in which the 
Boltzmann factors, $e^{-\beta_j E_j/2}$, are ignored and $j_{\rm max}$ 
is replaced by the effective Hilbert space dimension for the $\ket{b_j}$ 
states, which we identify as the Hilbert space dimension for the 
$\ket{\tilde{b}_j}$ states.  The conditions in Eq.~(\ref{eq:ortho-max}), 
therefore, must be regarded as approximate ones.

For simplicity, below we will use the expressions in Eqs.~(\ref{eq:psi_i},~%
\ref{eq:ortho-max}) for $\ket{\psi_i}$'s, which is a good approximation for 
our purposes.  The more precise expression of Eq.~(\ref{eq:psi_i-Boltzmann}), 
however, suggests why the number of independent black hole vacuum states 
$\ket{\psi_i}$ is only $e^{{\cal A}/4 l_{\rm P}^2}$, despite the fact 
that the dimension of the Hilbert space for the combined $\tilde{B}B$ 
system is much larger, ${\rm dim}\,{\cal H}_{\tilde{B}B} = e^{{\cal A}/2 
l_{\rm P}^2}$.  If maximal entanglement between $\tilde{B}$ and $B$ were 
the only condition for a smooth horizon, then we would have $e^{{\cal A}/2 
l_{\rm P}^2}$ smooth horizon black hole states.  In order for the horizon 
to be smooth, however, the $\tilde{B}$ and $B$ states must be entangled 
in a particular Boltzmann weighted way; in particular, $\ket{b_j}$ having 
energy $E_j$ must be multiplied by $\ket{\tilde{b}_j}$ having exactly the 
opposite energy $-E_j$, not by some $\ket{\tilde{b}_k}$ with $E_k \neq 
-E_j$.  Here, the concept of energy for the $\tilde{B}$ states arises through 
identification of these states as the modes outside the horizon in the 
other side of an eternal black hole; see Section~\ref{subsec:ext-int}. 
Assuming that there are no $B$ states exactly degenerate in energy, 
this only leaves a room to put phase factors in front of various 
$\ket{\tilde{b}_j} \ket{b_j}$ terms, leading to only ${\rm dim}\,{\cal H}_B$ 
(not ${\rm dim}\,{\cal H}_{\tilde{B}B}$) independent states as shown 
in Eq.~(\ref{eq:psi_i-Boltzmann}), where ${\rm dim}\,{\cal H}_B$ is the 
effective Hilbert space dimension for the $\ket{b_j}$ states.  Taking the 
number of independent black hole states to be $e^{{\cal A}/4 l_{\rm P}^2}$ 
as implied by the standard thermodynamic argument, the dimensions of 
${\cal H}_{\tilde{B}}$ and ${\cal H}_B$ are fixed as in Eq.~(\ref{eq:H_Bs}). 
As emphasized in Refs.~\cite{Verlinde:2013uja,Verlinde:2013vja}, this 
implies that space spanned by the states $\ket{\psi_i}$ comprises 
only a tiny, $e^{{\cal A}/4 l_{\rm P}^2}$-dimensional, subspace of 
the Hilbert space representing the combined $\tilde{B}B$ system: 
${\rm dim}\,{\cal H}_{\tilde{B}B} = e^{{\cal A}/2 l_{\rm P}^2} 
\gg e^{{\cal A}/4 l_{\rm P}^2}$.%
\footnote{Incidentally, the Bekenstein-Hawking entropy may also be 
 identified as the von~Neumann entropy of a reduced density matrix 
 $\rho_{CR}$ for the combined system $CR$ obtained after integrating 
 out the horizon degrees of freedom $\tilde{B}$, along the lines of 
 Ref.~\cite{Sorkin}.  For this identification to work, the fundamental 
 length scale $l_*$ must scale as $l_*^2 \sim N l_{\rm P}^2$, where 
 $N$ is the number of species appearing in low energy 4-dimensional 
 field theory applicable at length scales larger than $l_*$.  We assume 
 this is indeed the case (see e.g.~\cite{Dvali:2007hz}).}

In general, a black hole vacuum state can be represented by an arbitrary 
density matrix defined in space spanned by the $\ket{\psi_i}$'s.  In the 
case where entanglement between the black hole and the rest may be ignored, 
the entire system can be written as
\begin{equation}
  \ket{\Psi} \approx \left( \sum_{i=1}^{e^{{\cal A}/4 l_{\rm P}^2}} 
    c_i \ket{\psi_i} \right) \ket{r},
\label{eq:young}
\end{equation}
where $\ket{r}$ is an element of ${\cal H}_R$, the Hilbert space factor 
comprising all the possible states for subsystem $R$.  If the black 
hole is formed by a collapse of matter that has not been entangled 
with its environment, then the state of the system is well approximated 
by Eq.~(\ref{eq:young}) until later times (see below).  With such a 
formation, the number of possible black hole microstates is expected 
to be much smaller than $e^{{\cal A}/4 l_{\rm P}^2}$ (presumably of 
order $e^{c {\cal A}^{3/4}/l_{\rm P}^{3/2}}$ where $c$ is an $O(1)$ 
coefficient~\cite{'tHooft:1993gx}); but after the scrambling time 
$t_{\rm sc} \sim M_0 l_{\rm P}^2 \ln(M_0 l_{\rm P})$~\cite{Hayden:2007cs}, 
all these states are expected to evolve into {\it generic} states of 
the form in Eq.~(\ref{eq:young}):
\begin{equation}
  |c_i|^2 \sim O\Bigl( e^{-\frac{\cal A}{4 l_{\rm P}^2}} \Bigr).
\label{eq:generic}
\end{equation}
As time passes, the black hole becomes more and more entangled with 
the rest in the sense that the ratio of the entanglement entropy between 
$\tilde{B}B$ and $R$, $S_{\tilde{B}B} = S_R$, to the Bekenstein-Hawking 
entropy at that time, $S_{\rm BH} = 4\pi M^2 l_{\rm P}^2$, keeps growing, 
which saturates the maximum value $S_{\tilde{B}B} / S_{\rm BH} = 1$ 
after the Page time $t_{\rm Page} \sim M_0^3 l_{\rm P}^4$, where $M_0$ 
is the initial mass of the black hole~\cite{Page:1993wv}.  Therefore, 
the state of the system at late times must be written more explicitly 
as~\cite{Nomura:2012ex,Nomura,Nomura:2013nya}
\begin{equation}
  \ket{\Psi} = \sum_{i=1}^{e^{{\cal A}/4 l_{\rm P}^2}} 
    d_i \ket{\psi_i} \ket{r_i},
\label{eq:old}
\end{equation}
where $\ket{r_i}$'s are elements of ${\cal H}_R$.  In other words, 
at these late times the logarithm of the dimension of space spanned by 
$\ket{r_i}$'s is of order $S_{\rm BH}$ (and equal to $S_{\rm BH}$ after 
the Page time), while at much earlier times it is negligible compared 
with $S_{\rm BH}$.  The state at early times, therefore, can be well 
approximated by Eq.~(\ref{eq:young}) for the purpose of discussing 
internal properties of the black hole.

As we will see, the structure of the black hole states described above, 
together with dynamical assumptions discussed in Section~\ref{sec:basis}, 
elegantly addresses questions raised by the firewall argument.  Before 
turning to these issues, however, we make a comment on the structure 
of Hilbert space to avoid possible confusion.  As described at the 
beginning of this section, we have divided the system with a black hole 
{\it of mass $M$} into three subsystems $\tilde{B}$, $C$, and $R$; this 
division, therefore, implicitly depends on the mass $M$.  Since the 
black hole mass varies with time, the Hilbert space in which the state 
of the entire system evolves actually takes the form
\begin{equation}
  {\cal H} = \bigoplus_M \left( {\cal H}_{\tilde{B}(M)} \otimes 
    \left\{ {\cal H}_{B(M)} \oplus {\cal H}_{C(M)-B(M)} \right\} 
    \otimes {\cal H}_{R(M)} \right) 
  \equiv \bigoplus_M {\cal H}_M,
\label{eq:Hilbert}
\end{equation}
where we have explicitly shown the $M$ dependence of $\tilde{B}$, 
$B$, $C$, and $R$, and ${\rm dim}\,{\cal H}_{\tilde{B}(M)} = 
{\rm dim}\,{\cal H}_{B(M)} = e^{4\pi M^2 l_{\rm P}^2}$ as seen 
in Eq.~(\ref{eq:H_Bs}).  ${\cal H}_{C(M)-B(M)}$ is the Hilbert space 
spanned by the states for the $C$ degrees of freedom orthogonal to 
$B$ (i.e.\ not entangled with $\tilde{B}$), and we define ${\cal H}_0$ 
to be the Hilbert space for the system without a black hole.  As the 
black hole evolves, the state of the system moves between different 
${\cal H}_M$'s; for example, a state that is an element of ${\cal H}_{M_1}$ 
with some $M_1$ will later be an element of ${\cal H}_{M_2}$ with 
$M_2 < M_1$.%
\footnote{More precisely, the state at a later time will be a superposition 
 of elements in various ${\cal H}_{M_2}$'s, reflecting the probabilistic 
 nature of Hawking emission.  We ignore this effect, as well as a possible 
 spread of the initial black hole mass caused, e.g., by quantum effects 
 associated with the collapse, since they do not affect our argument.}

To help understand the meaning of Eq.~(\ref{eq:Hilbert}), let us consider 
a system in which a black hole was formed at time $t_0$ with the initial 
mass $M_0$: $\ket{\Psi(t_0)} \in {\cal H}_{M_0}$.  Suppose at some 
time $t$ with $t - t_0 \ll t_{\rm Page}$, the black hole mass is $M$ 
($< M_0$). Then, the state of the system at that time, $\ket{\Psi(t)}$, 
is given (approximately) by an element of ${\cal H}_M$ in the form of 
Eq.~(\ref{eq:young}).  Now, at a later time $t'$ with $t' - t_0 \simgt 
t_{\rm Page}$, the mass of the black hole becomes smaller, $M'$ ($< M$). 
The state of the system $\ket{\Psi(t')}$ is then given by an element 
of ${\cal H}_{M'}$ that takes the form of Eq.~(\ref{eq:old}) with 
(almost all) $\ket{r_i}$'s linearly independent.  Finally, after 
the black hole evaporates, the system is described by an element 
of ${\cal H}_0$ (which is generally time-dependent, representing 
the propagation of Hawking quanta).

\section{Black Hole Interior vs Firewalls}
\label{sec:basis}

In this section, we discuss the structure of elements in the Hilbert space 
factor ${\cal H}_{\tilde{B}} \otimes {\cal H}_B$ and operators acting on 
it, assuming that there is no extra matter near and outside the stretched 
horizon.  (If there is extra matter, it simply changes the identification 
of the $B$ states in the Hilbert space for the $C$ degrees of freedom.) 
For simplicity, we focus our discussion mostly on these entities for 
a fixed $M$.  The evolution of a black hole, which leads to a variation 
of $M$, is discussed in Section~\ref{subsec:evolution} only to the 
extent needed.

\subsection{Black hole states and firewall states}
\label{subsec:states}

Let the Hilbert space spanned by the black hole vacuum states $\ket{\psi_i}$ 
be ${\cal H}_\psi$:
\begin{equation}
  {\cal H}_\psi \subset {\cal H}_{\tilde{B}} \otimes {\cal H}_B.
\label{eq:H_psi}
\end{equation}
At the leading order, the dimension of ${\cal H}_\psi$ is 
$e^{{\cal A}/4 l_{\rm P}^2}$, i.e.
\begin{equation}
  \ln {\rm dim}\,{\cal H}_\psi = \frac{{\cal A}}{4 l_{\rm P}^2} 
    + O\left( \frac{{\cal A}^n}{l_{\rm P}^{2n}};\, n<1 \right) 
  \approx \frac{{\cal A}}{4 l_{\rm P}^2},
\label{eq:dim_H_psi}
\end{equation}
where ${\cal A} = 16\pi M^2 l_{\rm P}^4$.  (Here and below we use the 
approximate symbol, $\approx$, to indicate that an expression is valid at 
the leading order in expansion in inverse powers of ${\cal A}/l_{\rm P}^2$.) 
The basic idea of Refs.~\cite{Nomura:2012ex,Nomura} is that a semi-classical 
world can be constructed on each of $\ket{\psi_i}$'s, and that all of 
these worlds look identical to general relativity.  In the language here, 
this implies that an operator $\hat{O}$ that can be used to describe 
a semi-classical world for an infalling observer may be written in the 
block-diagonal form in the ${\cal H}_{\tilde{B}} \otimes {\cal H}_B$ space
\begin{equation}
  \hat{O} = \left(
  \begin{array}{cccc}
    \hat{o}_1 & \multicolumn{1}{:c}{} & \begin{array}{ccc}&&\\&&\\\end{array} 
    & \begin{array}{cc}&\\&\\\end{array} \\
    \cdashline{1-2}
    \begin{array}{cc}&\\&\\\end{array} & \multicolumn{1}{:c:}{\hat{o}_2} & 
    \multicolumn{2}{c}{\raisebox{2pt}[0pt][0pt]{\Huge $0$}} \\
    \cdashline{2-2}
    \begin{array}{cc}&\\&\\&\\\end{array} & 
    \begin{array}{cc}&\\&\\&\\\end{array} & 
    \ddots & 
    \begin{array}{cc}&\\&\\&\\\end{array} \\
    \cdashline{4-4}
    \begin{array}{cc}&\\&\end{array} &
    \raisebox{10pt}[0pt][0pt]{\Huge $0$} &
    \begin{array}{ccc} 
      &&\\&&\end{array} & 
    \multicolumn{1}{:c}{\hat{o}_{e^{\approx {\cal A}/4l_{\rm P}^2}}}
  \end{array} \right)
\!\!\!\!\!\!\!\!
  \begin{array}{l}
    \left. \begin{array}{c} \\ \\ \end{array} \right\} 
      e^{\approx \frac{\cal A}{4l_{\rm P}^2}} \\
    \left. \begin{array}{c} \\ \\ \end{array} \right\} 
      e^{\approx \frac{\cal A}{4l_{\rm P}^2}} \\
    \begin{array}{c} \\ \\ \\ \end{array} \qquad\vdots \\
    \left. \begin{array}{c} \\ \\ \end{array} \right\} 
      e^{\approx \frac{\cal A}{4l_{\rm P}^2}}
  \end{array}
\!\!\!\!\!\!\!\!\!
  \begin{array}{l}
    \left. \begin{array}{c} \\ \\ \\ \\ \\ \\ \\ \\ \\ \end{array} \right\} 
      e^{\approx \frac{\cal A}{2l_{\rm P}^2}} \\
  \end{array},
\label{eq:hat-O}
\end{equation}
if an appropriate basis is chosen.  Moreover, by taking an appropriate 
basis in each block, all the operators in the diagonal blocks $\hat{o}_i$, 
which we call {\it branch world operators} (and are represented here by 
$e^{\approx {\cal A}/4l_{\rm P}^2} \times e^{\approx {\cal A}/4l_{\rm P}^2}$ 
matrices), may be brought into an identical form
\begin{equation}
  \hat{o}_1 = \hat{o}_2 = \ldots 
    = \hat{o}_{e^{\approx {\cal A}/4l_{\rm P}^2}} \equiv \hat{o}.
\label{eq:hat-o}
\end{equation}
The resulting basis states can be arranged in the form
\begin{equation}
  \vec{e}_{\rm basis} 
  = \left(\!\!\! \begin{array}{c}
    \begin{array}{c} \vdots \\ \ket{\psi_1} \\\cdashline{1-1} \end{array} \\
    \begin{array}{c} \vdots \\ \ket{\psi_2} \\\cdashline{1-1} \end{array} \\
    \begin{array}{c} \\ \vdots\\ \\\end{array} \\
    \begin{array}{c} \cdashline{1-1} \vdots \\ 
      \ket{\psi_{e^{\approx {\cal A}/4l_{\rm P}^2}}} 
      \\\end{array}
  \end{array} \!\!\!\right),
\label{eq:basis}
\end{equation}
where we have listed the $e^{\approx {\cal A}/2l_{\rm P}^2}$ basis 
states in ${\cal H}_{\tilde{B}} \otimes {\cal H}_B$ in the form of a 
column vector; i.e., each block contains one of the black hole vacuum 
states $\ket{\psi_i}$'s, and in each block the $\ket{\psi_i}$ can be put 
at the bottom of the column vector of $e^{\approx {\cal A}/4l_{\rm P}^2}$ 
dimensions.

In the basis of Eq.~(\ref{eq:basis}), the outgoing creation/annihilation 
operators for an infalling observer ($\tilde{a}_{\tilde{\omega}}$ 
or $\hat{a}_{\omega}$ in Ref.~\cite{Unruh:1976db} or $a_\omega$ in 
Ref.~\cite{Almheiri:2012rt}) take the form of Eq.~(\ref{eq:hat-O}) 
with all the branch world operators taking the same form, which we 
denote by $\hat{a}_\omega$.%
\footnote{Strictly speaking, a creation/annihilation operator acting 
 on an element in ${\cal H}_M$ may transform it into an element of 
 ${\cal H}_{M'}$ with $M'$ (slightly) different from $M$.  Here we do 
 not treat this effect very explicitly, which can be done by considering 
 an appropriate set of ${\cal H}_M$'s (with appropriately coarse-grained 
 $M$'s, if one wants) or by defining the creation/annihilation operators 
 in such a way that when they act on an element in ${\cal H}_M$ the 
 resulting states stay in the same ${\cal H}_M$, by adjusting the mass 
 associated with the singularity at the center.  In any event, this 
 issue is not crucial for the firewall argument, since it can be 
 constructed using only number operators, e.g.\ $\hat{n}_{a_\omega} 
 = \hat{a}_\omega^\dagger \hat{a}_\omega$, which transform an element 
 of ${\cal H}_M$ into that in the same ${\cal H}_M$.}
(For simplicity, below we only consider spherically symmetric modes 
to keep the shape of the horizon, but the extension to other cases is 
straightforward.)  By acting (a finite number of) $\hat{a}_\omega^\dagger$'s 
on one of the $\ket{\psi_i}$'s, one can construct a state in which matter 
exists in the interior of the black hole as viewed from an infalling 
observer.  How many such states can we construct from a vacuum state 
$\ket{\psi_i}$, keeping the classical spacetime picture in the interior? 
We expect that the number of these states (for each $\ket{\psi_i}$) 
is of order $e^{\approx c {\cal A}^n / l_{\rm P}^{2n}}$ with 
$n < 1$~\cite{'tHooft:1993gx}.  This implies that the number of 
all the states in ${\cal H}_{\tilde{B}} \otimes {\cal H}_B$ that 
allow semi-classical interpretation in the black hole interior is
\begin{equation}
  e^{\approx \frac{\cal A}{4l_{\rm P}^2}} \times 
  e^{\approx c \frac{{\cal A}^n}{l_{\rm P}^{2n}}}
  = e^{\approx \frac{\cal A}{4l_{\rm P}^2}}.
\label{eq:n-classical}
\end{equation}
Namely, the Hilbert space factor ${\cal H}_{\rm cl}$ ($\supset 
{\cal H}_\psi$) spanned by all these semi-classical---i.e.\ not 
necessarily vacuum---black hole states satisfies
\begin{equation}
  \ln {\rm dim}\,{\cal H}_{\rm cl} 
  \approx \frac{{\cal A}}{4 l_{\rm P}^2},
\label{eq:H_cl-dim}
\end{equation}
consistent with the counting expected by the holographic 
bound~\cite{'tHooft:1993gx,Susskind:1994vu}.  As we will see more explicitly 
in the next subsection, an arbitrary superposition of elements of 
${\cal H}_{\rm cl}$ (or an arbitrary density matrix in ${\cal H}_{\rm cl} 
\otimes {\cal H}_{\rm cl}^*$) represents a black hole state in which 
an infalling observer sees smooth horizon.  In particular, this 
implies that in order for the infalling observer not to find any 
drama, the black hole state need not take the maximally entangled 
form in Eq.~(\ref{eq:psi_i})---it can even be in a separable form 
of $\ket{\tilde{b}_j} \ket{b_j}$ (without summation in $j$), since 
these states can be obtained as a superposition of (maximally 
entangled) $\ket{\psi_i}$'s.

Once again, the condition for an infalling observer to see smooth 
horizon is {\it not} that a black hole (vacuum) state has a maximally 
entangled form, but that it stays in the $e^{\approx {\cal A}/4l_{\rm 
P}^2}$-dimensional subspace ${\cal H}_{\rm cl}$ (or ${\cal H}_\psi$) 
in the $e^{\approx {\cal A}/2l_{\rm P}^2}$-dimensional space 
${\cal H}_{\tilde{B}} \otimes {\cal H}_B$ (called the balanced form 
in Ref.~\cite{Verlinde:2013uja} for the vacuum states).  This is 
because as long as the black hole state stays in ${\cal H}_{\rm cl}$, 
the dynamics of the horizon makes the observer see the state in the 
basis determined by $\ket{\psi_i}$, as will be discussed more explicitly 
in the next subsection.  This therefore replaces/refines (in a sense) 
the maximally-entangled condition of Ref.~\cite{VanRaamsdonk:2010pw} 
for the existence of smooth classical spacetime (in the present context, 
beyond the horizon). The vast majority of the states in ${\cal H}_{\tilde{B}} 
\otimes {\cal H}_B$ that do not belong to ${\cal H}_{\rm cl}$ are 
``firewall states.''  They do not admit the smooth classical spacetime 
picture in the interior of the horizon; in particular, they include 
states in which a diverging number of $a_\omega$ quanta, including 
high energy modes, are excited on $\ket{\psi_i}$.  It may be possible to 
view these states as representing the situation in which singularities of 
general relativity exist near the horizon, not just at the center (see 
Ref.~\cite{Susskind:2012rm}), so that there is no classical spacetime 
in the interior region.

\subsection{No firewall for black hole states---dynamical selection 
 of the basis}
\label{subsec:no-firewall}

We now argue that {\it if} the state stays in subspace ${\cal H}_{\rm cl}$ 
in the Hilbert space factor ${\cal H}_{\tilde{B}} \otimes {\cal H}_B$, 
then an infalling observer does not see firewalls.  We begin by discussing 
why operators responsible for describing (semi-)classical worlds 
for an infalling observer take the special block-diagonal form of 
Eqs.~(\ref{eq:hat-O},~\ref{eq:hat-o}).  Why can't general Hermitian 
operators acting on the $e^{\approx {\cal A}/2l_{\rm P}^2}$-dimensional 
space ${\cal H}_{\tilde{B}} \otimes {\cal H}_B$ be observables in these 
worlds?

As discussed in Refs.~\cite{Nomura:2011rb,q-Darwinism}, observables 
in classical worlds---which emerge dynamically from the full quantum 
dynamics---correspond, in general, to only a tiny subset of all the 
possible quantum operators acting on the microscopic state of the system. 
These observables represent the information that can be amplified in 
a single term in a quantum state (i.e.\ the information that can be 
shared and compared by multiple physical ``observers'' in the system), 
and are selected as a result of the dynamics of the system (the 
selection of the measurement basis).  The statement that operators 
used to describe a semi-classical world for an infalling observer 
take the form of Eqs.~(\ref{eq:hat-O},~\ref{eq:hat-o}), therefore, 
comprises an assumption on the {\it internal dynamics} of the 
$\tilde{B}B$ system, i.e.\ the microscopic Hamiltonian acting 
on the $\tilde{B}$ and $B$ degrees of freedom, which determines 
the form of operators representing observables in a semi-classical 
world~\cite{Nomura:2012ex}.  In fact, this is precisely the physical 
content of the complementarity hypothesis, which has to do with how 
classical spacetime emerges in a full quantum theory of gravity. 
In the distant description, an object falling into the horizon will 
interact strongly with surrounding highly blue-shifted Hawking quanta, 
making it (re-)entangled with the basis determined by the $\ket{\psi_i}$'s. 
Here we take this dynamical assumption for granted, which one might 
hope to eventually derive from the microscopic theory of the $\tilde{B}$ 
and $B$ degrees of freedom.

With this interpretation of semi-classical observables, it is now easy 
to see that an arbitrary state in ${\cal H}_{\rm cl}$, or more 
generally an arbitrary density matrix in ${\cal H}_{\rm cl} \otimes 
{\cal H}_{\rm cl}^*$, does not lead to a firewall for an infalling 
observer.  Consider, for simplicity, that the $\tilde{B}$ and $B$ degrees 
of freedom are in a pure state
\begin{equation}
  \ket{\psi} = \sum_{i=1}^{e^{\approx {\cal A}/4 l_{\rm P}^2}}\!\! 
    c_i \ket{\psi_i},
\label{eq:BH-pure}
\end{equation}
where $c_i$'s are arbitrary coefficients with $\sum_{i=1}^{e^{\approx 
{\cal A}/4 l_{\rm P}^2}} |c_i|^2 = 1$.  If the infalling observer interacts 
with this state, then he/she will ``measure,'' or ``feel,'' it in the 
basis determined by $\hat{O}$'s in Eqs.~(\ref{eq:hat-O},~\ref{eq:hat-o}); 
i.e.\ he/she will find that the black hole is in a particular state 
$\ket{\psi_i}$ with probability $|c_i|^2$.  Since all the $\ket{\psi_i}$ 
states represent the same semi-classical black hole with smooth horizon 
(at the level of general relativity), this implies that the observer 
will find that the horizon is smooth with a probability of $1$---the 
observer does not see a firewall.%
\footnote{This statement is exact if the measurement basis is selected 
 perfectly by Eqs.~(\ref{eq:hat-O},~\ref{eq:hat-o}), which will indeed 
 be the case in the limit that the $\tilde{B}B$ system has an infinite 
 number of degrees of freedom.  Since the number of $\tilde{B}B$ degrees 
 of freedom is very large, though not infinite, this is an extremely 
 good approximation---the correction, if any, will be exponentially 
 suppressed by a factor of $e^{-O(S_{\rm BH})}$.}

In fact, one can obtain the same conclusion by calculating the average 
number of high energy $a_\omega$ quanta (i.e.\ $a_\omega$ quanta with 
$\omega \gg 1/M l_{\rm P}^2$) for the states in ${\cal H}_\psi$.  In 
the basis of Eq.~(\ref{eq:basis}), the number operators for $a_\omega$ 
modes take the form
\begin{equation}
  \hat{N}_{a_\omega} = \left. \left(
  \begin{array}{cccc}
    \hat{n}_{a_\omega} & \multicolumn{1}{:c}{} & 
    \begin{array}{ccc}&&\\&&\\\end{array} & 
    \begin{array}{cc}&\\&\\\end{array} \\
    \cdashline{1-2}
    \begin{array}{cc}&\\&\\\end{array} & 
    \multicolumn{1}{:c:}{\hat{n}_{a_\omega} } & 
    \multicolumn{2}{c}{\raisebox{2pt}[0pt][0pt]{\Huge $0$}} \\
    \cdashline{2-2}
    \begin{array}{cc}&\\&\\&\\\end{array} & 
    \begin{array}{cc}&\\&\\&\\\end{array} & 
    \ddots & 
    \begin{array}{cc}&\\&\\&\\\end{array} \\
    \cdashline{4-4}
    \begin{array}{cc}&\\&\end{array} &
    \raisebox{10pt}[0pt][0pt]{\Huge $0$} &
    \begin{array}{ccc}&&\\&&\end{array} & 
    \multicolumn{1}{:c}{\hat{n}_{a_\omega}}
  \end{array} \right)
  \right\} \, e^{\approx \frac{\cal A}{2l_{\rm P}^2}},
\qquad
  \hat{n}_{a_\omega} = \hat{a}_\omega^\dagger \hat{a}_\omega
  \approx \left. 
    \left( \begin{array}{cccc}
      &&& 0 \\ &&& \vdots\\ & 
      \multicolumn{2}{l}{\raisebox{6pt}[0pt][0pt]{\,\,\,\,\Huge *}} & \vdots\\
      0 & \cdots & \cdots & 0
    \end{array} \right)
    \right\} \, e^{\approx \frac{\cal A}{4l_{\rm P}^2}},
\label{eq:n_a}
\end{equation}
for all $\omega \gg 1/M l_{\rm P}^2$, since $\ket{\psi_i}$'s are black 
hole vacuum states.  The average number of high energy $a_\omega$ quanta 
is then
\begin{equation}
  \bar{N}_{a_\omega} \equiv \frac{{\rm Tr}_{{\cal H}_\psi} 
    \hat{N}_{a_\omega}}{{\rm Tr}_{{\cal H}_\psi} {\bf 1}} 
  = \frac{\sum_{i=1}^{e^{\approx {\cal A}/4 l_{\rm P}^2}} \bra{\psi_i} 
    \hat{N}_{a_\omega} \ket{\psi_i}}{e^{\approx {\cal A}/4 l_{\rm P}^2}} 
  \approx 0,
\label{eq:average}
\end{equation}
for any $\omega \gg 1/M l_{\rm P}^2$, where the traces are taken over 
arbitrary basis states in ${\cal H}_\psi$.  (Note that $\bar{N}_{a_\omega}$ 
is independent of the basis chosen.)  Since an expectation value of a 
number operator, $\hat{N}_{a_\omega}$, is positive semi-definite, this 
implies that typical states in ${\cal H}_\psi$ do {\it not} have firewalls. 
By the definition of ${\cal H}_{\rm cl}$, the same argument also applies 
to the states in ${\cal H}_{\rm cl}$.

In Ref.~\cite{Marolf:2013dba}, a similar calculation was performed with 
the conclusion that typical black hole states {\it do} have firewalls. 
A crucial element in the calculation of Ref.~\cite{Marolf:2013dba} was 
the statement/assumption that the eigenstates of the number operator 
$\hat{b}^\dagger \hat{b}$ provide a complete basis for unentangled 
black hole states (see also~\cite{Bousso}), where $\hat{b}$ is the 
annihilation operator for a Killing mode that is located outside 
the stretched horizon.  If this were true, then we could calculate 
the average number of high energy quanta for the black hole states 
$\bar{N}_{a_\omega}$, defined analogously to Eq.~(\ref{eq:average}), 
by going to the basis spanned by the $\hat{b}^\dagger \hat{b}$ 
eigenstates.  Now, the semi-classical relation
\begin{equation}
  \hat{b} = \int\! d\omega \left( \beta(\omega)\hat{a}_\omega 
    + \gamma(\omega)\hat{a}_\omega^\dagger \right),
\label{eq:b_vs_a}
\end{equation}
where $\beta(\omega)$ and $\gamma(\omega)$ are some functions, implies 
that the expectation value of an $a_\omega$-number operator in a 
$\hat{b}^\dagger \hat{b}$ eigenstate is $O(1)$ for any $\omega \gg 
1/M l_{\rm P}^2$.  This would, therefore, give $\bar{N}_{a_\omega} 
\approx O(1)$, implying that typical black hole states must have 
firewalls.

Why has our calculation led to the opposite conclusion?  The key point 
is that with the structure of the Hilbert space discussed here, the 
traces over ${\cal H}_\psi$ in Eq.~(\ref{eq:average}) cannot be taken 
as those over $\hat{b}^\dagger \hat{b}$ eigenstates.  Below, we examine 
this point more closely.  While doing so, we also discuss the structure 
of quantum operators that can be used to describe the exterior and 
interior spacetime regions of the black hole.

\subsection{Exterior and interior operators}
\label{subsec:ext-int}

As we have seen in Section~\ref{subsec:states}, the creation/annihilation 
operators for quanta on $\ket{\psi_i}$'s take the form of 
Eqs.~(\ref{eq:hat-O},~\ref{eq:hat-o}) (e.g.\ with $\hat{o} = 
\hat{a}_\omega$ and $\hat{a}_\omega^\dagger$ for outgoing modes), 
which generically act nontrivially both on the $\tilde{B}$ and $B$ 
degrees of freedom.  In general, one may take a linear combination 
of these operators to construct operators that represent a mode localized 
in the exterior or interior of the horizon (with the latter viewed 
from an infalling observer).  What form would such operators take?

Let us consider the annihilation operator $\hat{b}$ for an outgoing 
mode that is localized outside the stretched horizon.  We consider a mode 
in $B$, i.e.\ a mode (significantly) entangled with stretched horizon 
degrees of freedom $\tilde{B}$.%
\footnote{In view of Eq.~(\ref{eq:psi_i-Boltzmann}), such a mode has 
 an energy of the order of, or smaller than, the local Hawking temperature. 
 Modes that have significantly higher energies than local Hawking 
 temperatures are not directly entangled with $\tilde{B}$.}
This is the mode used in the argument of Ref.~\cite{Marolf:2013dba}.
The operator $\hat{b}$ can be constructed by taking a linear combination 
of creation/annihilation operators $\hat{O}$'s with $\hat{o} = 
\hat{a}_\omega$ and $\hat{a}_\omega^\dagger$.  Because of the assumption 
that low energy physics outside the stretched horizon is well described 
by local quantum field theory built on $C$ ($\supset B$) and $R$ degrees 
of freedom, this operator must take the form
\begin{equation}
  \hat{b} = {\bf 1} \otimes \hat{b}_B \qquad 
    \mbox{in } {\cal H}_{\tilde{B}} \otimes {\cal H}_B,
\label{eq:hat-b}
\end{equation}
i.e.\ act only on the $B$ degrees of freedom, where $\hat{b}$ and 
$\hat{b}_B$ are operators defined in $e^{\approx {\cal A}/2 l_{\rm 
P}^2}$-dimensional and $e^{\approx {\cal A}/4 l_{\rm P}^2}$-dimensional 
Hilbert spaces, respectively.  The complementarity hypothesis asserts 
that semi-classical physics---in particular physics responsible for 
the Hawking radiation process---persists, implying that the relation 
in Eq.~(\ref{eq:b_vs_a}) must be preserved (with $\hat{a}_\omega$ and 
$\hat{a}_\omega^\dagger$ interpreted as the corresponding operators 
in the full $e^{\approx {\cal A}/2 l_{\rm P}^2}$-dimensional Hilbert 
space).  This implies that the action of the particular linear 
combination of $\hat{a}_\omega$'s and $\hat{a}_\omega^\dagger$'s 
appearing in the right-hand side of Eq.~(\ref{eq:b_vs_a}) on $\tilde{B}$ 
must be trivial, so that $\hat{b}$ takes the form of Eq.~(\ref{eq:hat-b}).

What about operators representing an interior mode?  One might naively 
think that those operators, collectively written as $\hat{d}$, take 
the form $\hat{d} = \hat{d}_{\tilde{B}} \otimes {\bf 1}$, analogous 
to Eq.~(\ref{eq:hat-b}).  However, the structure of the black hole 
vacuum states in Eq.~(\ref{eq:psi_i}) (or Eq.~(\ref{eq:psi_i-Boltzmann})) 
and that of the Hilbert space in Eq.~(\ref{eq:H_Bs}) suggest that this 
is not the case.  These structures are exactly those of near-horizon 
modes of an eternal black hole with the same mass $M$.  We therefore 
postulate that
\begin{itemize}
\item[]
The {\it quantum mechanical} structure of a black hole after the 
scrambling time (when formed by a collapse) is the same as that of 
an eternal black hole {\it (even) at the microscopic level}.
\end{itemize}
In particular, the relevant Hilbert space describing a black hole 
of mass $M$ has dimension $e^{\approx {\cal A}/2 l_{\rm P}^2}$ 
($= e^{\approx 8\pi M^2 l_{\rm P}^2}$), not $e^{\approx {\cal A}/4 
l_{\rm P}^2}$, although the dynamics will make the state sweep only 
$e^{\approx {\cal A}/4 l_{\rm P}^2}$-dimensional subspace of it as 
the time passes or as the initial condition for the collapse is scanned. 
(This sweeping will be ergodic in the subspace after a sufficient 
coarse-graining.)  This is, obviously, consistent with the semi-classical 
expectation that a black hole formed by a collapse looks like an eternal 
black hole when it is probed late enough.  Here we require that it is 
also the case quantum mechanically at the microscopic level, including 
the form of operators representing various excitations.

More precisely, we consider that the $B$ and $\tilde{B}$ degrees of freedom 
for a black hole of mass $M$ correspond, respectively, to the near-horizon 
degrees of freedom in one and the other external regions---often called 
regions I and III---of an eternal black hole {\it with the same mass $M$} 
as viewed from a distant reference frame.  (See Fig.~\ref{fig:map} for a 
schematic depiction.)  Here, the near-horizon modes are defined such that 
the reactions of the modes in region~I to the exterior operators of the 
form in Eq.~(\ref{eq:hat-b}) are the same as those of $B$; for example, 
these modes have energies of the order of, or smaller than, local Hawking 
temperatures.  The near-horizon modes in region~III can then be 
defined through entanglement with those in region~I.  We assume 
that the Hilbert space structure of the $B$ and $\tilde{B}$ states 
for a collapse-formed black hole (often called a one-sided black 
hole) is the same as that of the states in the two exterior regions 
of an eternal black hole (two-sided black hole) with the quantization 
hypersurface taken as an equal-time hypersurface determined by the 
outside timelike Killing vector.  This suggests that operators responsible 
for describing the interior spacetime region take the form
\begin{equation}
  \hat{d} = \hat{d}_{\tilde{B}} \otimes {\bf 1} 
    + {\bf 1} \otimes \hat{d}_B \qquad 
  \mbox{in } {\cal H}_{\tilde{B}} \otimes {\cal H}_B,
\label{eq:hat-d}
\end{equation}
where $\hat{d}$ is defined in the full $e^{\approx {\cal A}/2 l_{\rm 
P}^2}$-dimensional Hilbert space ${\cal H}_{\tilde{B}} \otimes {\cal H}_B$, 
while $\hat{d}_{\tilde{B}}$ and $\hat{d}_B$ are defined in $e^{\approx 
{\cal A}/4 l_{\rm P}^2}$-dimensional Hilbert spaces ${\cal H}_{\tilde{B}}$ 
and ${\cal H}_B$, respectively.

The structure of operators in Eq.~(\ref{eq:hat-d}) can be motivated 
by the fact that an equal-time hypersurface determined by the outside 
timelike Killing vector forms a Cauchy surface, so that all the local 
operators in the interior spacetime can be obtained by evolving local 
operators of the form $\hat{o}_{\rm region\,\,III} \otimes {\bf 1} + {\bf 1} 
\otimes \hat{o}_{\rm region\,\,I}$ on the initial equal-time hypersurface 
along a set of hypersurfaces depicted by dotted lines in the right 
panel of Fig.~\ref{fig:map}.  (Note that this evolution preserves this 
particular form of operators, since it can be represented by acting 
some unitary operator $U$ and its inverse $U^{-1}$ from left and 
right, respectively.)  We assume that the interior region of the 
black hole under consideration can be constructed in this way with 
$\hat{o}_{\rm region\,\,I}$ and $\hat{o}_{\rm region\,\,III}$ acting 
only on the near-horizon modes in the respective regions, leading to 
Eq.~(\ref{eq:hat-d}).  We emphasize again that the ``identification'' 
of the $B$ and $\tilde{B}$ states with the eternal black hole states 
is made at each instant of time (or in a sufficiently short time period 
compared with the timescale for the evolution of the black hole); in 
particular, the mass of the eternal black hole must be taken as that 
of the evolving black hole at each moment $M(t)$, not the initial 
mass $M_0$.  This implies that an infalling object passes through 
the horizon of the eternal black hole at the center of the Penrose 
diagram depicted in the right panel of Fig.~\ref{fig:map}.

We are now at the position of discussing why our analysis in 
Section~\ref{subsec:no-firewall} has led to the opposite conclusion 
as that in Ref.~\cite{Marolf:2013dba}.  The key point is that in our 
present framework, the black hole vacuum states $\ket{\psi_i}$ provide 
a complete basis of the ($e^{\approx {\cal A}/4 l_{\rm P}^2}$-dimensional) 
Hilbert space ${\cal H}_\psi$ ($\subset {\cal H}_{\tilde{B}} \otimes 
{\cal H}_B$), while the $\hat{b}^\dagger \hat{b}$ eigenstates 
provide that of a {\it different} (again, $e^{\approx {\cal A}/4 
l_{\rm P}^2}$-dimensional) Hilbert space ${\cal H}_B$.  This implies, 
in particular, that the black hole vacuum states $\ket{\psi_i}$ can 
{\it not} all be made $\hat{b}^\dagger \hat{b}$ eigenstates by performing 
a unitary rotation in the space spanned by $\ket{\psi_i}$ (i.e.\ 
${\cal H}_\psi$) in which the traces in Eq.~(\ref{eq:average}) are 
taken.  This can be seen more explicitly as follows.  Let us write 
$\ket{\psi_i}$'s in the form of Eq.~(\ref{eq:psi_i}).  Since the 
$\hat{b}^\dagger \hat{b}$ eigenstates span a basis in ${\cal H}_B$, 
we may write $\ket{b_j}$ as
\begin{equation}
  \ket{b_j} = \sum_{k=1}^{e^{\approx {\cal A}/4l_{\rm P}^2}}\!\! 
    c^j_k \ket{e_k},
\label{eq:bb-expand}
\end{equation}
where $\ket{e_k}$'s are the $\hat{b}^\dagger \hat{b}$ eigenstates. 
(Note that these eigenstates may be degenerate; i.e., $\ket{e_k}$ 
and $\ket{e_{k'}}$ with $k \neq k'$ need not have different eigenvalues 
under $\hat{b}^\dagger \hat{b}$.)  Substituting Eq.~(\ref{eq:bb-expand}) 
into Eq.~(\ref{eq:psi_i}), we obtain
\begin{equation}
  \ket{\psi_i} = \sum_{k=1}^{e^{\approx {\cal A}/4 l_{\rm P}^2}} 
    \left( \sum_{j=1}^{e^{\approx {\cal A}/4 l_{\rm P}^2}}
    \alpha^{(i)}_j c^j_k \ket{\tilde{b}_j} \right) \ket{e_k} 
  \equiv \sum_{k=1}^{e^{\approx {\cal A}/4 l_{\rm P}^2}}\! 
    f^{(i)}_k \ket{\tilde{e}^{(i)}_k} \ket{e_k},
\label{eq:psi_i-exp}
\end{equation}
where
\begin{equation}
  \ket{\tilde{e}^{(i)}_k} 
  \propto \sum_{j=1}^{e^{\approx {\cal A}/4 l_{\rm P}^2}}\! 
    \alpha^{(i)}_j c^j_k \ket{\tilde{b}_j}.
\label{eq:tilde-e_k}
\end{equation}
An important point is that the element in ${\cal H}_{\tilde{B}}$ that 
is entangled with $\ket{e_k}$, i.e.\ $\ket{\tilde{e}^{(i)}_k}$, depends 
on the vacuum state, i.e.\ on the index $i$.  This prevents us from finding 
a basis change in ${\cal H}_\psi$ that makes all the $\ket{\psi_i}$'s 
$\hat{b}^\dagger \hat{b}$ eigenstates.  (Otherwise, the rotation represented 
by the matrix $(f^{-1})_{(i)}^k$ would do the job.)  The traces in 
Eq.~(\ref{eq:average}), therefore, cannot be taken over $\hat{b}^\dagger 
\hat{b}$ eigenstates, avoiding the conclusion in Ref.~\cite{Marolf:2013dba}.

\subsection{Dynamical evolution}
\label{subsec:evolution}

We have seen that the $\tilde{B}$ and $B$ degrees of freedom can represent 
very different objects---black holes and firewalls---depending on their 
quantum mechanical states.  In particular, if they are in a state that 
is an element of $e^{\approx {\cal A}/4 l_{\rm P}^2}$-dimensional Hilbert 
space ${\cal H}_{{\rm cl}(M)}$ (or represented by a density matrix that 
is an element of ${\cal H}_{{\rm cl}(M)} \otimes {\cal H}_{{\rm cl}(M)}^* 
\equiv {\cal L}_{{\rm cl}(M)}$), then an infalling observer interacting 
with these degrees of freedom will find that the horizon is smooth and see 
the interior spacetime; if not, he/she will see a firewall.  (In general, 
if the $\tilde{B}B$ state involves a component in ${\cal H}_{{\rm cl}(M)}$, 
the infalling observer will see the interior spacetime with the 
corresponding probability.)  Here, we have restored the index $M$ 
to remind us that ${\cal H}_{{\rm cl}(M)}$ is, in fact, a component 
of ${\cal H}_M$; see Eq.~(\ref{eq:Hilbert}).

A natural interpretation of these black hole and firewall states---i.e.\ 
the elements of ${\cal H}_M$---is that they both represent objects that 
lead to the Schwarzschild spacetime of mass $M$ in the region outside 
the (stretched) horizon, because they both use the same $C$ ($\supset B$) 
and $R$ degrees of freedom and local quantum field theory is supposed 
to be valid in this region.  A crucial question then is:\ what does the 
dynamics of the (entire) system tell us about the properties of an object 
under consideration?  In particular, if the object is formed by a collapse 
of matter with the initial mass $M_0$, are the corresponding $\tilde{B}$ 
and $B$ degrees of freedom in a state in ${\cal H}_{{\rm cl}(M_0)}$ 
(or ${\cal L}_{{\rm cl}(M_0)}$)?  And if so, do they stay in 
${\cal H}_{{\rm cl}(M)}$ (or ${\cal L}_{{\rm cl}(M)}$) when 
the mass is reduced to $M$ ($< M_0$) by time evolution?

We do not know {\it a priori} the answer to these questions.  We may, 
however, interpret the success of general relativity with the global 
spacetime picture to mean that the dynamical evolution keeps the $\tilde{B}B$ 
degrees of freedom to stay in subspace ${\cal H}_{{\rm cl}(M)}$ (or 
${\cal L}_{{\rm cl}(M)}$), i.e.\ in a state that allows for the smooth 
classical spacetime interpretation in the interior of the horizon (at 
least, in the absence of a certain special manipulation of the degrees 
of freedom by an infalling observer; see Section~\ref{sec:infalling}). 
This interpretation can be made particularly plausible by considering 
the extension of the framework to (meta-stable) de~Sitter space, where 
the validity of the global spacetime picture beyond the de~Sitter horizon 
is strongly supported by the successful prediction for density perturbations 
in the inflationary universe; see Section~\ref{sec:de-Sitter}.  We 
therefore postulate that the dynamics of quantum gravity is such that 
it keeps an element of ${\cal H}_{{\rm cl}(M)}$ (or ${\cal L}_{{\rm cl}(M)}$) 
to stay in ${\cal H}_{{\rm cl}(M')}$ (or ${\cal L}_{{\rm cl}(M')}$) under 
time evolution, where $M' \neq M$ in general.  In the context of black 
hole physics, this implies that the state of the system takes the form of 
Eq.~(\ref{eq:young}) when the black hole is formed by (isolated) matter, 
which evolves into Eq.~(\ref{eq:old}) as time passes.  This evolution 
is indeed consistent with the standard dynamics for the black hole 
evaporation process, e.g.\ the generalized second law of thermodynamics. 
An explicit qubit model representing this process was described in 
Ref.~\cite{Verlinde:2013vja}.

As discussed in Ref.~\cite{Nomura:2012ex}, the evolution described 
above is also consistent with the standard analysis of the information 
flow in black hole evaporation in a unitary theory of quantum 
gravity~\cite{Page:1993wv}, avoiding the paradox raised by 
Ref.~\cite{Almheiri:2012rt}.  (In fact, this was how the structure 
of the black hole states discussed in this paper was first found.) 
In Ref.~\cite{Almheiri:2012rt}, it was argued that for an old black 
hole, the conditions for unitarity and smooth horizon, which were 
respectively given by
\begin{equation}
  S_{BR} < S_R,
\qquad
  S_{\tilde{B}B} \approx 0,
\label{eq:AMPS}
\end{equation}
were mutually incompatible, where $S_X$ represents the von~Neumann entropy 
of subsystem $X$.  The structure of the black hole states discussed here, 
however, elegantly avoids this conclusion, keeping the standard assumptions 
of black hole complementarity.  For a state with an old black hole given 
in Eq.~(\ref{eq:old}), the conditions for unitarity and smooth horizon 
are given, respectively, by~\cite{Nomura:2012ex}
\begin{equation}
  S_{BR} < S_R,
\qquad
  \tilde{S}^{(i)}_{\tilde{B}B} \approx 0 \;\; (\mbox{for all }i),
\label{eq:NV}
\end{equation}
where $\tilde{S}^{(i)}_X$ are {\it branch world entropies}, the von~Neumann 
entropy of subsystem $X$ calculated using the state representing 
the (semi-)classical world $i$: $\ket{\Psi^{(i)}} = \ket{\psi_i}\ket{r_i}$ 
(without summation in the right-hand side).  The point is that the 
relations in Eq.~(\ref{eq:NV}) are {\it not} incompatible with each 
other; in fact, they are all satisfied for a generic state of the form 
of Eq.~(\ref{eq:old}) after the Page time.  The reason for this success 
can be put as follows:
\begin{itemize}
\item[]
What is responsible for unitarity of the evolution of the black hole 
state is not an entanglement between early radiation and modes in $B$ 
as imagined in Ref.~\cite{Almheiri:2012rt}, but an entanglement between 
early radiation and {\it the way $B$ and $\tilde{B}$ degrees of freedom 
are entangled}.
\end{itemize}
In other words, the information of the black hole is contained in the 
coefficients $d_i$ in Eq.~(\ref{eq:old}), i.e.\ how the black hole is 
made out of the $e^{\approx {\cal A}/4 l_{\rm P}^2}$ vacuum states.

\subsection{Gauge/gravity duality}
\label{subsec:dual}

Here we comment on how a black hole (or a firewall) may be realized in 
the gauge theory side of gauge/gravity duality.  As an example, we might 
consider a setup in Ref.~\cite{Almheiri:2013hfa} in which an evaporating 
black hole is modeled by a conformal field theory (CFT) coupled to a 
large external system.  How can the structure discussed in this paper 
be realized in such a setup?

One possibility is that the whole structure of the Hilbert space described so 
far, in particular the Hilbert space of $e^{\approx {\cal A}/2 l_{\rm P}^2}$ 
dimensions for the horizon and the entangled degrees of freedom, 
exists at the microscopic level in the corresponding gauge theory. 
The whole $e^{\approx {\cal A}/2 l_{\rm P}^2}$ degrees of freedom 
are not visible in standard thermodynamic considerations in the gauge 
theory side, since the dynamics populates only the $e^{\approx {\cal A}/4 
l_{\rm P}^2}$ subspace (${\cal H}_{\rm cl}$) of the whole Hilbert space 
(${\cal H}_{\tilde{B}} \otimes {\cal H}_B$) relevant for these degrees 
of freedom.  The local operators responsible for describing the exterior 
and interior regions are still given in the form of Eqs.~(\ref{eq:hat-b}) 
and (\ref{eq:hat-d}), respectively.  We note that a similar construction 
of operators in the exterior and interior regions were discussed in 
Ref.~\cite{Papadodimas:2012aq}, in which the $\tilde{B}$ degrees of 
freedom were thought to arise effectively after coarse-graining the 
outside degrees of freedom.  Our picture here is different---we consider 
that both the stretched horizon ($\tilde{B}$) and the outside ($B$) degrees 
of freedom exist independently at the microscopic level.  It is simply 
that standard black hole thermodynamics does not probe all the degrees 
of freedom because of the properties of the dynamics.

An alternative possibility is that the gauge theory only contains states 
in ${\cal H}_{\rm cl}$ for the horizon and the entangled degrees of freedom.%
\footnote{We thank Donald Marolf for discussions on this point.}
In this case, the gauge theory cannot describe a process in which the 
state for these degrees of freedom is made outside ${\cal H}_{\rm cl}$ 
(i.e.\ creation of a firewall), or it may simply be that such a process 
does not exist in the gravity side as well (see a related discussion 
in Section~\ref{subsec:inf-obs}).

\section{Infalling Observer}
\label{sec:infalling}

In this section, we discuss what our framework predicts for the fate of 
infalling observers under various circumstances.  We also discuss how 
the physics can be described in an infalling reference frame, rather 
than in a distant reference frame as we have been considering so far.

\subsection{Physics of an infalling observer: a black hole or firewall?}
\label{subsec:inf-obs}

As we have seen, as long as the black hole state is represented by 
a density matrix in ${\cal L}_{\rm cl} = {\cal H}_{\rm cl} \otimes 
{\cal H}_{\rm cl}^*$, an infalling observer interacting with this state 
sees a smooth horizon with a probability of $1$.  Here we ask what 
happens if the observer manipulates the relevant degrees of freedom, 
e.g.\ measures some of the $B$ or $R$ modes, before entering the 
horizon.  This amounts to asking what black hole state such an observer 
encounters when he/she falls into the horizon.

We first see that the observer finds a smooth horizon with probability 
$1$ no matter what measurement he/she performs on Hawking radiation 
emitted earlier from the black hole.  This statement is obvious if the 
object the observer measures is not entangled with the black hole, e.g.\ 
as in Eq.~(\ref{eq:young}), so we consider the case in which the observer 
measures an object that is entangled with the black hole.  To be specific, 
we consider the case in which the observer measures the $\ket{r_i}$ 
degrees of freedom in Eq.~(\ref{eq:old}) before he/she enters the 
black hole.  Without loss of generality, we assume that the outcome 
of the measurement was $\sum_k U_{jk} \ket{r_k}$, where $U_{jk}$ is 
an arbitrary unitary matrix.  The state of the black hole the observer 
encounters is then $\sum_k U^\dagger_{kj} d_k \ket{\psi_k}$.  Since 
this is a superposition of the black hole states $\ket{\psi_i}$'s, the 
observer finds a smooth horizon with a probability of $1$, as discussed 
in Section~\ref{subsec:no-firewall}.  We conclude that it is not possible 
to create a firewall by making a measurement on early Hawking radiation.

On the other hand, if an infalling observer can directly access the $B$ 
states entangled with the stretched horizon modes, then he/she may be able 
to see a firewall.  Consider that the black hole state $\ket{\Psi}$ is 
given by Eq.~(\ref{eq:old}).  We may then expand the $B$ states in terms 
of the $\hat{b}^\dagger \hat{b}$ eigenstates as in Eq.~(\ref{eq:bb-expand}), 
leading to
\begin{equation}
  \ket{\Psi} = \sum_{i,k=1}^{e^{\approx {\cal A}/4 l_{\rm P}^2}}\! 
    d_i f^{(i)}_k \ket{\tilde{e}^{(i)}_k} \ket{e_k} \ket{r_i},
\label{eq:firewall-1}
\end{equation}
where $f^{(i)}_k$ and $\ket{\tilde{e}^{(i)}_k}$ are given in 
Eqs.~(\ref{eq:psi_i-exp},~\ref{eq:tilde-e_k}).  Suppose the observer 
measures the $B$ degrees of freedom are in a $\hat{b}^\dagger \hat{b}$ 
eigenstate $\ket{e_k}$.  Then the relevant state is
\begin{equation}
  \ket{\Psi} \propto \sum_{i=1}^{e^{\approx {\cal A}/4 l_{\rm P}^2}}\! 
    d_i f^{(i)}_k \ket{\tilde{e}^{(i)}_k} \ket{e_k} \ket{r_i},
\label{eq:firewall-2}
\end{equation}
without summation in $k$.  Now, the state $\ket{r_i}$ is in general 
a superposition of decohered classical states $\ket{r^{\rm cl}_j}$, 
$\ket{r_i} = \sum_j g^{(i)}_j \ket{r^{\rm cl}_j}$,%
\footnote{The reason for this decomposition is that decohered terms 
 in a state must be treated as different worlds when calculating 
 probabilities.  Note that the way $\ket{r_i}$ is decomposed into 
 $\ket{r^{\rm cl}_j}$'s is not important for our discussion (which 
 simply alters the values of $g^{(i)}_j$), i.e.\ our conclusion 
 is not affected by the way this decoherence happens.}
so the observer finds the $\tilde{B}B$ state is in one of the 
$\ket{\tilde{\psi}_j}$'s, where
\begin{equation}
  \ket{\tilde{\psi}_j} \propto 
    \sum_{i=1}^{e^{\approx {\cal A}/4 l_{\rm P}^2}}\! 
    d_i f^{(i)}_k g^{(i)}_j \ket{\tilde{e}^{(i)}_k} \ket{e_k}.
\label{eq:firewall-3}
\end{equation}
This state is {\it not} in general a superposition of the smooth horizon 
states $\ket{\psi_i}$'s; in other words, $\ket{\tilde{\psi}_j}$ is not 
an element of ${\cal H}_{\rm cl}$.  Thus, if the infalling observer 
directly measures $B$ to be in a $\hat{b}^\dagger \hat{b}$ eigenstate 
(which will require the detector to be finely tuned) and enters the 
horizon right after (i.e.\ before the state of the black hole changes), 
then he/she will see a firewall with an $O(1)$ probability.%
\footnote{There is a small probability that the observer still sees 
 a smooth horizon, since a superposition of different $k$ states in 
 Eq.~(\ref{eq:firewall-2}) in which the firewall terms are canceled must 
 be able to reproduce the smooth horizon state, Eq.~(\ref{eq:firewall-1}).}
Here we have assumed that such a measurement can be performed, 
although it is possible that there is some dynamical (or perhaps 
computational~\cite{Harlow:2013tf}) obstacle to it.  We note that 
a similar argument to the one here applies to any surface in a low 
curvature region, not just the black hole horizon.  This way of seeing 
a firewall, therefore, does not violate the equivalence principle.

We finally mention that even if an observer finds $B$ to be in a 
$\hat{b}^\dagger \hat{b}$ eigenstate, if he/she enters the horizon 
long after that, then he/she may see a smooth horizon rather than 
a firewall.  Suppose that the observer measured a $\hat{b}^\dagger 
\hat{b}$ eigenstate when the black hole had mass $M$.  This implies 
that he/she was entangled with the $\hat{b}^\dagger \hat{b}$ eigenstate 
in $B(M)$, where we have explicitly shown the $M$ dependence of the 
decomposition.  Now, at a much later time, the black hole has a smaller 
mass $M'$ ($\ll M$).  The mode which was in $B(M)$ may then be in 
$R(M')$.  If this is the case, then the observer is no longer entangled 
with a $B$ mode necessary to see a firewall, i.e.\ $B(M')$; instead, 
he/she is simply entangled with the environment (early radiation) of 
the black hole, $R(M')$.  This observer will therefore see a smooth 
horizon when he/she enters the black hole, as discussed at the beginning 
of this subsection.  The fact that an observer was once entangled with 
a $B$ mode is not enough to see a firewall; he/she must be entangled 
with a $B$ mode of the black hole at the time of entering the horizon 
in order to see a firewall.

\subsection{Description in an infalling reference frame}
\label{subsec:inf-desc}

We now consider how the physics for an infalling object is described 
in an infalling reference frame, rather than in a distant frame as has 
been considered so far.  Following Ref.~\cite{Nomura:2011rb}, we consider 
that such an infalling description is obtained by performing a unitary 
transformation on the distant description, corresponding to a change 
of the clock degrees of freedom in full quantum gravity.  According to 
the complementarity hypothesis, the Hamiltonian---the generator of 
time evolution---after the transformation takes the form local in 
infalling field operators, including the interior operators discussed 
in Section~\ref{subsec:ext-int}.

Since the complementarity transformation is supposed to be unitary, 
the $e^{\approx {\cal A}/4 l_{\rm P}^2}$ states $\ket{\psi_i}$ must be 
transformed into $e^{\approx {\cal A}/4 l_{\rm P}^2}$ different states 
which must all look like locally Minkowski vacuum states.  In particular, 
this implies that in the limit that the black hole is large ${\cal A} 
\rightarrow \infty$, i.e.\ in the limit that the horizon under consideration 
is a Rindler horizon, there are infinitely many Minkowski vacuum states 
labeled by $i=1,\cdots,e^{\approx {\cal A}/4 l_{\rm P}^2}=\infty$.  This 
seems to contradict our experience that we can do physics without knowing 
which of the Minkowski vacua we live in.  Isn't the Minkowski vacuum 
unique, e.g., in QED?  Otherwise, we do not seem to be able to do any 
physics without having the (infinite amount of) information on the 
Minkowski vacua.

The structure of operators discussed in Section~\ref{subsec:states}, 
however, provides the answer.  After the complementarity transformation, 
the form of operators is preserved; in particular, all the local operators 
responsible for describing semi-classical physics take the block-diagonal 
form, Eq.~(\ref{eq:hat-O}), with all the (branch world) operators in 
the diagonal blocks taking the identical form, Eq.~(\ref{eq:hat-o}). 
This implies that no matter which superposition of Minkowski vacua 
we live in, we always find the same semi-classical physics (which is 
everything in the non-gravitational limit).  More precisely, if we live 
in a vacuum represented by state $c_i \ket{\psi_i}$ ($\sum_i |c_i|^2 = 1$), 
then we find ourselves to be in a particular vacuum $\ket{\psi_i}$ with 
probability $|c_i|^2$ (i.e.\ the measurement basis is $\ket{\psi_i}$), 
but all of these vacua lead to the same semi-classical physics.  In 
order to discriminate different vacua, we need to consider operators 
beyond local field operators, e.g.\ those of Eq.~(\ref{eq:hat-O}) with 
$\hat{o}_i$ taking different forms for different $i$ in the basis of 
Eq.~(\ref{eq:basis}).  Such operators will be either highly nonlocal 
or act on the boundary/horizon of the infalling/locally Minkowski 
description of spacetime.  In the true Minkowski space, the boundary 
is located only at spatial infinity.  In the case of an infalling 
description of a black hole vacuum, however, spacetime is Minkowski 
vacuum-like only locally, and the nonzero curvature effect can lead 
to a horizon (as viewed from the infalling frame, not the original one 
as viewed from a distant frame) at a finite spatial distance, e.g.\ 
along the lines of Ref.~\cite{Nomura:2013nya}.  Probing microscopic 
degrees of freedom on such a horizon, therefore, might allow us to access 
the information on which $\ket{\psi_i}$ vacuum the observer is in.

\section{de~Sitter Space}
\label{sec:de-Sitter}

The quantum theory of horizons described in this paper is applicable 
to horizons other than those of black holes with straightforward 
adaptations.  Here we discuss an application to de~Sitter horizons. 
The analysis is parallel with the case of black hole horizons.

The de~Sitter horizon is located at $r = 1/H$ in de~Sitter space, where 
$r$ is the radial coordinate of the static coordinate system $(t, r, 
\theta, \phi)$ and $H$ the Hubble parameter.  The stretched horizon 
is located where the local Hawking temperature
\begin{equation}
  T(r) = \frac{H/2\pi}{\sqrt{1-H^2 r^2}},
\label{eq:T_dS}
\end{equation}
becomes of order the fundamental/string scale $M_* = 1/l_*$: $T(r_*) 
= M_*/2\pi$ (the factor of $2\pi$ is for convenience), i.e.\
\begin{equation}
  r_* = \frac{1}{H} - \frac{H}{2M_*^2}.
\label{eq:dS-stretched}
\end{equation}
Following the analysis of the black hole case, we consider that the 
dimensions of the Hilbert spaces for the stretched horizon degrees of 
freedom, $\tilde{B}$, and the states entangled with them, $B$, are 
given by
\begin{equation}
  {\rm dim}\,{\cal H}_{\tilde{B}} = {\rm dim}\,{\cal H}_B 
  = e^{\frac{\cal A}{4 l_{\rm P}^2}},
\label{eq:H_dS}
\end{equation}
where ${\cal A} = 4\pi/H^2$ is the area of the (stretched) de~Sitter 
horizon.  We may expect that at the leading order in $l_{\rm P}^2/{\cal A}$, 
the dimension of the Hilbert space spanned by all the possible states 
inside the de~Sitter horizon ($r < r_*$) is the same as that of 
${\cal H}_B$, which we assume to be the case.  (As can be seen from 
the distribution of thermal entropy $\propto T(r)^3$, most of these 
states are localized near the horizon.)  The dimension of the total 
Hilbert space needed to describe de~Sitter space is then
\begin{equation}
  {\rm dim}\,({\cal H}_{\tilde{B}} \otimes {\cal H}_B) 
  = e^{\frac{\cal A}{2 l_{\rm P}^2}},
\label{eq:dim_dS}
\end{equation}
at the leading order in $l_{\rm P}^2/{\cal A}$.

As in the case of black holes, the states which allow for the interpretation 
of classical spacetime outside the de~Sitter horizon span only a tiny 
subspace of the entire Hilbert space
\begin{equation}
  {\cal H}_{\rm cl} \subset {\cal H}_{\tilde{B}} \otimes {\cal H}_B,
\qquad
  {\rm dim}\,{\cal H}_{\rm cl} = e^{\frac{\cal A}{4 l_{\rm P}^2}}.
\label{eq:dS-cl}
\end{equation}
In fact, the Hilbert space spanned by the vacuum states already has the 
same logarithmic dimension at the leading order in $l_{\rm P}^2/{\cal A}$:
\begin{equation}
  {\cal H}_\psi \subset {\cal H}_{\rm cl},
\qquad
  {\rm dim}\,{\cal H}_\psi = e^{\frac{\cal A}{4 l_{\rm P}^2}},
\label{eq:dS-psi}
\end{equation}
with the basis states for measurement taking the maximally entangled form 
as in Eq.~(\ref{eq:psi_i}) (ignoring Boltzmann factors).  As long as a 
state stays in ${\cal H}_{\rm cl}$ (or is represented by a density matrix 
in ${\cal H}_{\rm cl} \otimes {\cal H}_{\rm cl}^*$), an object that hits 
the horizon can be thought of going to space outside the horizon; otherwise, 
it hits a firewall/singularity.  The information about the object that 
goes outside will be stored in the $\tilde{B}B$ degrees of freedom, 
which may later be recovered, for example if the system evolves into 
Minkowski space (or another de~Sitter space with a smaller vacuum 
energy).  As in the black hole case, we expect that the dynamics of 
quantum gravity is such that a state in ${\cal H}_{\rm cl}$ keeps staying 
in ${\cal H}_{\rm cl}$.  In fact, this is observationally indicated by 
the success of the prediction for density perturbations in the inflationary 
universe, which is based on the global spacetime picture of general 
relativity~\cite{Hawking:1982cz}.

Finally, we consider the limit $H \rightarrow 0$ in which the de~Sitter 
space approaches Minkowski space.  In this limit, the number of vacuum 
states becomes infinity
\begin{equation}
  {\rm dim}\,{\cal H}_\psi \rightarrow \infty,
\label{eq:H-Minkoski}
\end{equation}
each differing in the way $\tilde{B}$ and $B$ states are entangled; see 
Eq.~(\ref{eq:psi_i}).  Since most of the $B$ states are localized near 
the horizon, which is located at spatial infinity in this limit, probing 
the structure of (infinitely many) Minkowski vacua will require access 
to the boundary at infinity.  This is the same picture we have arrived 
at in Section~\ref{subsec:inf-desc} by taking the large mass limit of 
a black hole horizon.

The fact that various horizons, especially black hole and de~Sitter 
horizons, can be treated on an equal footing is an important ingredient 
for the quantum mechanical treatment of the eternally inflating 
multiverse advocated in Refs.~\cite{Nomura:2011dt,Nomura:2011rb}, 
in which the eternally inflating multiverse and the many worlds 
interpretation of quantum mechanics are unified as the same concept. 
It would be interesting to see if there are other implications of 
the present framework beyond what have been discussed in this paper.

\section*{Acknowledgments}

We thank Juan Maldacena for useful conversations.  This work was supported 
in part by the Director, Office of Science, Office of High Energy 
and Nuclear Physics, of the US Department of Energy under Contract 
DE-AC02-05CH11231, and in part by the National Science Foundation 
under grant PHY-1214644.

\end{document}